# Convective Instability Of The Solar Corona: Why The Solar Wind Blows

Joseph Lemaire

*Institut d'Aéronomie Spatiale de Belgique (3, Avenue circulaire, B-1080 Bruxelles, Belgium)*
*jfl@astr.ucl.ac.be*

**Abstract.** Chapman's[10] conductive model of the solar corona is characterized by a temperature varying as $r^{-2/7}$ with heliocentric distance $r$. The density distribution in this non-isothermal hydrostatic model has a minimum value at 123 $R_S$, and increases with $r$ above that altitude. It is shown that this hydrostatic model becomes convectively unstable above $r=35\ R_S$, where the temperature lapse rate becomes superadiabatic. Beyond this radial distance heat conduction fails to be efficient enough to keep the temperature gradient smaller than the adiabatic lapse rate. We report the results obtained by Lemaire[12] who showed that an additional mechanism is then required to transport the energy flux away from the Sun into interplanetary space. He pointed out that this additional mechanism is advection: i.e. the stationary hydrodynamic expansion of the corona. In other words the corona is unable to stay in hydrostatic equilibrium. The hydrodynamic solar wind expansion is thus a physical consequence of the too steep (superadiabatic) temperature gradient beyond the peak of coronal temperature that can be determined from white light brightness distributions observed during solar eclipses. The thermodynamic argument for the existence of a continuous solar wind expansion which is presented here, complements Parker's classical argument based on boundary conditions imposed to the solutions of the hydrodynamic equations for the coronal expansion: i.e. the inability of the mechanical forces to hold the corona in hydrostatic equilibrium. The thermodynamic argument presented here is based on the energy transport equation. It relies on the temperature distribution which becomes super-adiabatic above a certain altitude in the inner corona.

## INTRODUCTION

The physical origin of the extended bright halo of light surrounding the Sun during solar eclipses has remained mystery for many decades. The discovery of its physical origin is generally attributed to Edlén[1]. Indeed, from Edlén's pioneering spectroscopic identification of the yellow and green coronal emission lines it became clear that the solar coronal gas is formed of highly ionized atoms. This coronal plasma was assumed to trapped within the Sun's gravitational, magnetic and electric fields [2].

It is generally not known that Alfvén[3], had already published one year earlier, in the same Scandinavian Journal a paper suggesting that the corona was an extremely hot atmosphere formed of fully ionized Hydrogen ions and electrons. He inferred this result from his correct interpretation of the small slope of the coronal brightness as a function of heliocentric distances. Considering that the corona is an extended atmosphere of fully ionized Hydrogen ions and of free electrons, he determined from the calculated density scale height that the coronal temperature must have a peak value of 1.98 $10^6$ K, between 1.2 and 3 solar radii. The coronal temperature determined by Alfvén in 1941, decreases on both sides of this maximum, as illustrated by Fig. 1 taken from Alfvén's [3].

Besides the early suggestion by Vandt[2] that this hot coronal gas should be formed by "appreciable evaporation of finely divided interplanetary matter [accreted] in the vicinity of the Sun", there have been many other heating mechanisms proposed to maintain the solar corona at such high temperatures; theyare reported in many papers and textbooks [4,5,6].

The aim of the present article is to use theoretical and empirical temperature distributions, like that illustrated in Fig.1, to demonstrate that the solar corona is convectively unstable, and that it must necessarily expand as inferred in 1958 by Parker[7]. The thermodynamic argument developed below, is based on the energy transport equation in the corona; it complements Parker's one which is based boundary conditions imposed on his hydrodynamic solutions of the steady state coronal expansion.

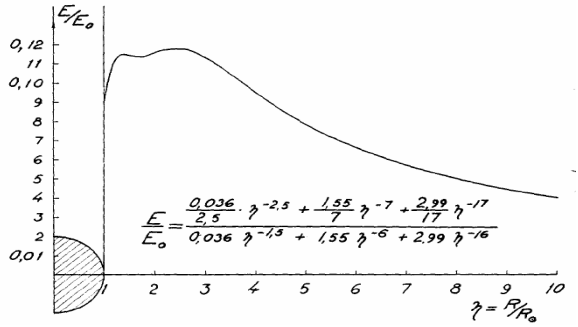

Figure 1. This plot shows the first radial distribution of the coronal temperature as determined in 1941, by Alfvén[3] from the scale height of an empirical radial density profile of the coronal electron inferred from eclipse observations. The coronal temperature $T(R)$ is related to $E(R)$ by $E=(3/2)kT$. The coronal temperature has a maximum between 1.5 and 3 $R_S$ : $T_{max} = 1.98 \times 10^6$ K $= 0.12$ $E_o$, where $E_o = (3/4) g_o R_S m_H = 1490$ eV.

## HYDROSTATIC MODELS OF THE SOLAR CORONA

### Empirical Coronal Temperature and Density Distributions.

Using the same method as Alfvén[3], Pottasch[8] calculated $T(r)$, a similar radial coronal temperature distribution from the density scale heights of $n_e(r)$, another equatorial density distribution deduced, from observations of the 1952 eclipse. The solid curve with symbols in Fig. 2 shows $T(h)$ as a function $h$, the altitude between $h=1.2R_S$, and $20R_S$ [1]. Empirical coronal temperature profiles derived from more contemporary eclipse observations show similar trends. Smaller peak values are found over the poles.

Pottasch's temperature profile was calculated under the assumptions: (i) that the corona is in hydrostatic equilibrium, (ii) that its brightness and density distributions are spherically symmetric, (iii) that it is formed of 90 % $H^+$ ions and 10% $He^{++}$ ions, (iv) that the ion and electron temperatures are the same, and finally (v) that the plasma is quasi-neutral. The peak temperature is $T_{max} = 1.43 \; 10^6$ K at $h = 0.5R_S$. Note that Brandt[9] showed that $T(h)$ is not significantly affected for $h < 3R_S$, when a subsonic hydrodynamic expansion of the inner corona is assumed, instead of hydrostatic equilibrium.

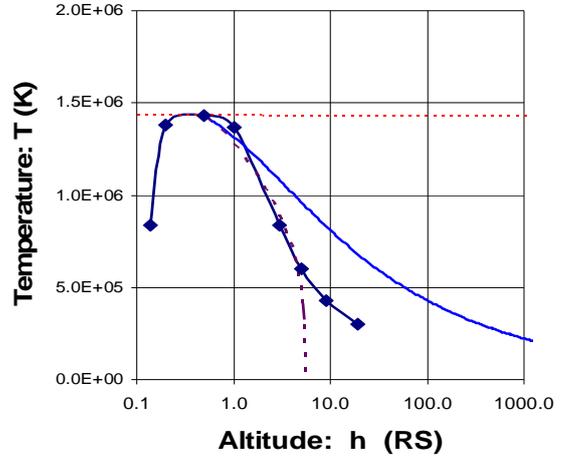

Figure 2. Coronal temperatures versus altitude above the photosphere. The diamond symbols: Pottasch's[8] empirical temperature distribution deduced from eclipse observations; peak value: $T_{max} = 1.43 \times 10^6$ K. The solid line at $h > 0.5$ $R_S$ is Chapman's[10] conductive temperature model. The dotted line is an isothermal coronal model similar to that of van de Hulst[11]. The dotted-dashed line is a conductive temperature distribution fitting Pottasch's empirical temperature gradient somewhere between $h = 2$ $R_S$ and $4$ $R_S$.

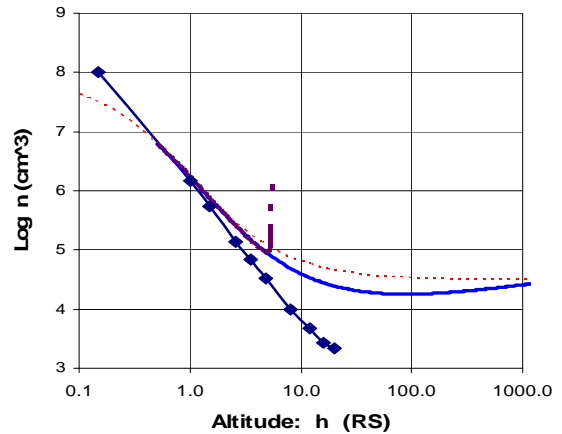

Figure 3. Coronal electron densities versus the altitude above the photosphere. Diamond symbols show Pottasch[8] empirical densities. Solid thick line is Chapman's[10] conductive hydrostatic model. Dotted line: isothermal hydrostatic model for $T(h) = 1.43 \times 10^6$ K. The doted-dashed line with steep vertical slope at $h = 5.2$ $R_S$ is a density obtained with an alternative conductive hydrostatic model fitted to match Pottasch's[8] temperature gradient somewhere between $h = 2$ $R_S$ and $4$ $R_S$.

---

[1] There is definite advantage to use the altitude $h$ instead of the radial distance $r$, to map the distributions of density, temperature, scale height, etc…in the inner corona, where these physical quantities vary drastically over small distances.

In Fig. 2 the solid line extending from $h = 0.5\ R_S$ to $1000 R_S$ corresponds to Chapman's[10] temperature profile; it varies as $r^{-2/7}$; it is his solution of the thermal conduction equation for which $T(h) = 0$ at $h = \infty$.

Using this analytical temperature profile in the hydrostatic equation one obtains the density distribution illustrated by the solid line in Fig.3. The dotted line gives the electron density profile if the coronal would be isothermal [$T(h) = 1.43 \times 10^6\ K$], and in hydrostatic equilibrium like those generally considered before 1957 by van de Hulst[11] or others.

The curve with the symbols corresponds to Pottasch[8] density profile employed to derive the empirical coronal temperature profile shown in Fig. 1. It can be seen that for $r < 4\ R_S$, white light eclipse observations are rather well fitted by hydrostatic density profiles, whether the coronal temperature distribution is isothermal or given by Chapman's conductive solution. This indicates that coronal density distributions are not very sensitive to the *gradient of the coronal temperature*, while they are very sensitive to the maximum value of *T(h)*. This property was not pointed out in earlier studies. It could be exploited to determine from available white light coronal brightness observations, reliable 'peak values' for coronal plasma temperatures at different heliographic latitudes.

At infinity, the isothermal density and pressure distributions tend to values which are much larger than the values expected in the interstellar medium. This led Parker[7] to conclude in 1958, that the kinetic temperature and pressure at the base of the corona are too high to sustain hydrostatic equilibrium up to infinity. The excessive pressure predicted by hydrostatic models at large distances, prompted him to argue that the corona must continuously expand, forming what he called then the "Solar Wind" (SW).

## Arguments In Support Of The Hydrodynamic Expansion Of The Corona

Parker's argument for a continuous supersonic expansion of the corona is a mechanical argument. It is based on boundary conditions imposed at infinity and at the base of the corona, for the stationary solutions of the mass and momentum transport hydrodynamic equations in their simplest formulation: the Euler approximation.

As can be seen from Fig. 3, the situation is even worse than for the density profile in Chapman's hydrostatic model. Indeed, in this case the radial density distribution has a minimum at $123\ R_S$, and beyond that altitude the density increases beyond any limit... Of course, any atmosphere where the density distribution increases with altitude is necessarily convectively unstable. Eddy convection is then expected to avoid denser plasma elements to reside above the less dense ones.

The physical origin for this unstable theoretical trend was pointed out by Lemaire[12]. He noted that this unstable trend is the consequence of Chapman's temperature profile becoming super-adiabatic already at $r = 34\ R_S$. This implies then that in Chapman's conductive coronal temperature model the temperature gradient exceeds the marginal value corresponding to Schwarzschild's convective stability criterion. This means that energy transport by heat conduction alone is not efficient enough to carry the heat away from the coronal temperature peak into interplanetary space. A similar situation occurs in the Earth's troposphere, where the temperature becomes equal to the adiabatic lapse rate; eddy convection is then setting up to carry more efficiently energy upwards into the stratosphere[2].

The same situation is thus recovered in the corona where some additional energy transport mechanism is also required to help evacuating more efficiently than by heat conduction all the energy deposited at its base.

A similar process operates also in the Hydrogen convection zone (HCZ) of the Sun, below the photosphere. Using a mixing-length theory, like that developed by Vitense[13] for the HCZ and others [14] for planetary atmospheres, Lemaire[12] developed a coronal model taking into account eddy convection in the unstable layers of Chapman's corona: hot plasma elements boiling upward, and the cooled one buoying down in a corona still in hydrostatic equilibrium in the average, i.e. for turbulent energy transport, but with no net transport of mass.

Lemaire's calculations showed that additional turbulent energy transport is not adequate to prevent the temperature gradient to stay below the marginal threshold in a conductive-convective model of the solar corona; indeed, in Lemaire's mixing length model the turbulent eddies must achieve supersonic velocities to maintain the temperature gradient less steep than the super-adiabatic lapse rate.

In other words, even within such a HCZ-type coronal model, it is not possible to evacuate efficiently enough the energy flux flowing upward away from the peak of temperature.

This is how Lemaire[12] was led to the conclusion that only an hydrodynamic expansion (i.e. a continuous expansion or explosion) of the coronal plasma is capable to evacuate most efficiently the energy with all the way through into outer space. This can be viewed as a new thermodynamic argument in support of a solar wind expansion of the solar corona.

---

[2] Let's remind here that a large fraction of the energy deposited in the inner corona is also conducted inwards by heat conduction into the chromosphere, as a consequence of the steep positive temperature gradient in the transition region[15].

## Thermodynamic Argument For A Continuous Expansion Of The Solar Corona

Thus it is reasonable to view the coronal plasma acceleration as 'advection of plasma' driven by a too steep temperature gradient in the inner corona produced by a heat source like a furnace or a 'stationary explosion'.

The new argument for the existence of stationary advection instead of hydrostatic equilibrium in the corona, is a supplementary physical argument in support of the solar wind expansion. In addition to Parker's mechanical argument for an hydrodynamic expansion of the corona, the present thermodynamic one is based on the energy transport equation determined in the inner corona, while Parker's original argument was based on unbalanced pressure (mechanical) forces, and on a discussion on boundary conditions of the various solutions of the mass and momentum transport equations. The new thermodynamic argument presented here, reinforces the proposition of Parker[7] which is based on the Euler formulation of the hydrodynamic equations.

Since some are claiming, however, that the Euler approximation of the Hierarchy of Moment Equations fails to be applicable above a certain exobase level, the same people might argue that it is questionable to ground ones argument on solutions of equations whose validity is only limited to the collision dominated region of the corona.

## More Deterrent Solutions Of The Heat Conduction Equation

Besides Chapman's special solution of the heat conduction equation for which the temperature tends to zero at infinity, there are other solutions with even steeper temperature gradients. The dashed curves in Figs. 2 and 3 illustrate one of them whose temperature gradient matches Pottasch's empirical temperature distribution, between $h = 2$ and $4 R_S$. This temperature profile becomes already super-adiabatic (i.e. convectively unstable) at $h = 2.3 R_S$, the hydrostatic density distribution has a minimum value at $h = 4.6 R_S$ and a very steep positive gradient higher up.

This leads to the conclusion that hydrostatic equilibrium is becoming convectively unstable, even deeper in the solar corona than in Chapman's popular conductive model, and that the expansion of the corona is initiated quite near its base.

## CONCLUSION

This presentation has lead to definitely disregard hydrostatic models of the solar corona, and to adopt a continuous expansion like that described by hydrodynamic and kinetic models of the solar wind. We can expect that this conclusion should also stand for other stellar coronae to that of our Sun.

## ACKNOWLEDGMENTS

I dedicate this article to late Prof. P. Ledoux, ULg. The preparation of this contribution as well as my participation at the SW12 conference has been supported by the Belgian Institute for Space Aeronomy and the LOC of the SW12 conference. I wish to thank N. Meyer-Vernet for her invitation, and for her standing interest in the kinetic theory of space plasmas.